\documentstyle[12pt]{article}
\newcommand{\beq}{\begin{equation}}
\newcommand{\eeq}{\end{equation}}
\newcommand{\bea}{\begin{eqnarray}}
\newcommand{\eea}{\end{eqnarray}}
\newcommand{\bay}{\begin{array}}
\newcommand{\eay}{\end{array}}
\newcommand{\hb}{h_{v}^{(b)}}                 %bottom-field
\newcommand{\hbbar}{\bar h_{v}^{(b)}}         %bottom bar-field

\newcommand{\rslash}{\mbox{$\not{\hspace{-1.03mm}r}$}}        % rslash
\newcommand{\qslash}{\mbox{$\not{\hspace{-1.03mm}q}$}}        % rslash
\newcommand{\vslash}{\mbox{$\not{\hspace{-1.03mm}v}$}}        % vslash
\newcommand{\dslash}{\mbox{$\not{\hspace{-1.03mm}D}$}}        % Dslash
\newcommand{\vcross}{\mbox{$\times{\hspace{-2.5mm}v}$}}      % vcross
\begin{document}
\begin{titlepage}
\begin{flushleft}
MZ-TH/94-03\\
February 1994
\end{flushleft}
\begin{center}
\large
\bf
{\Large Heavy quark distribution functions from QCD}\\[2cm]
\rm
D.Pirjol$^1$\\[.5cm]
Johannes Gutenberg-Universit\"at\\
Institut f\"ur Physik (THEP), Staudingerweg 7\\
D-55099 Mainz, Germany\\[2cm]
\normalsize
\bf
Abstract\\
\rm
\small
\end{center}
The structure function of a heavy quark in a heavy hadron in electroproduction
is calculated from QCD. Its deviation from the parton model prediction (a delta
function) is shown to be governed by a shape function, similar to the one
recently proposed by Neubert and Bigi {\em et al.} to describe the end-point
behaviour of the lepton spectrum in semileptonic $B\to X_ue\bar\nu$ decays
and the line shape in $B\to X_s\gamma$. Two new sum rules are derived which
extend the Bjorken-Dunietz-Taron sum rule by including corrections suppressed
by the inverse heavy quark mass.
\\
[2cm]
\footnotesize
$^1\,$Supported by the Graduiertenkolleg Teilchenphysik, Universit\"at Mainz\\
\normalsize
\end{titlepage}

 {\bf 1.} There have been recently significant advances in the theoretical
understanding of the inclusive decays of hadrons containing one heavy
quark. It has been shown that a number of model--independent predictions can
be obtained, for example for the lepton energy spectrum in semileptonic
decays of B mesons or the photon energy spectrum in the decay $B\to X_s
\gamma$ \cite{6,FLS}.
The formal tool which was used is an adaptation of the operator product
expansion method (OPE) to deal with this specific situation \cite{CGG},
in which the large mass scale (the heavy quark mass) is transferred from
the matrix elements of the operators to the Wilson coefficients.
{}From an intuitive point of view, the results thus obtained can be
interpreted as corrections to the naive free--quark decay predictions,
due to a ``Fermi motion'' of the heavy quark inside the hadron.
Initially introduced by Neubert to describe the end--point region
behaviour of the lepton spectrum in $B\to X_ue\bar\nu$ \cite{M1}, the
concept of a
``shape function'' turned out to be a process--independent parametrization
of the ``Fermi motion'', incorporating features which cannot be reproduced
by a quantum--mechanical description in terms of a wavefunction
\cite{M2,F,R}.

  In this paper we discuss the $b$ quark structure functions in
electroproduction on a B meson and show that they can be calculated in
QCD into an expansion in powers of $\bar\Lambda/m_b$. We give explicit
results for these function to second order in this parameter. To leading
order in this expansion the parton model prediction is obtained, which
is a simple delta function. The corrections to it have a singular form
consisting of delta functions and its derivatives. It is shown
that the most singular contributions can be summed up to all orders in
$1/m_b$ into a ``shape function'' which extends beyond the parton model
region and which can be considered a good approximation to the real
distribution function.

   In Section 4, two generalizations of the Bjorken sum rule are derived
\cite{BJ}. The sum rules result
from equating the total contribution to the moments of the heavy quark
structure functions in electroproduction on a heavy meson (as obtained by
the OPE analysis) to the sum of the individual contributions of the
resonances. Both these quantities can be arranged as an expansion in
$1/m_Q$, the inverse heavy quark mass, and at the leading order in this
expansion the usual Bjorken sum rule is recovered. At higher orders in
$1/m_Q$ the resulting sum rules are new and can be used to set constraints
on the nonperturbative subleading form--factors.

  The spirit of the approach and the nature of the approximations which
are made are related to those recently used \cite{6} to calculate the
inclusive semileptonic decays of heavy hadrons containing a heavy quark.
However, the formalism used here is more similar to the one familiar from
deep--inelastic scattering. Instead of modifying the operator-product
expansion (OPE) so that the heavy--mass dependence resides from beginning
in the coefficient functions, we will prefer to perform the OPE in the full
theory (QCD), and to expand the resulting matrix elements in powers of
$1/m_Q$. The equivalence of the two approaches is proved on an explicit
example to order $1/m_Q^2$.

 The heavy quark structure functions in electroproduction on a heavy meson
have been considered previously by Jaffe and Randall \cite{JR}. In this paper
we go further by giving explicit expressions for quantities which were
just parametrized in \cite{JR}. Also, some of the results described in
Section 3 have been obtained by Neubert \cite{M2} in a light-cone language.
Our derivation is perhaps more direct as it proceeds directly from the
usual definitions of the structure functions.\\[0.5cm]

  {\bf 2.} Let us consider the forward Compton scattering matrix element
on a heavy meson $B$
\beq
T_{\mu\nu}(q\cdot p, Q^2) = -i\int\mbox{d}^4\! xe^{-iq\cdot x}
 \langle B(v)|T(\bar b\gamma_\mu b)(x) (\bar b\gamma_\nu b)(0)|B(v)\rangle\,,
\eeq
where $p=m_B v$ is the B--meson momentum and $Q^2=-q^2>0$ is the photon
virtuality. The meson state $|B(v)\rangle$ has the usual normalization
$2m_Bv_0$. This situation is similar to what one usually has in
deep-inelastic scattering, in that the virtual photon is spacelike. However,
in distinction to the latter, we will not require that $Q^2 \gg m_B^2$,
but will take $Q^2, m_B^2 \gg \Lambda_{QCD}$ (with the exception of Section 3,
where we take temporarily $Q^2\to\infty$). We will keep therefore all
contribution of order $m_B^2/Q^2$ and
organize the result as an expansion in powers of $\Lambda/m_B$.

  $T_{\mu\nu}$ can be decomposed into covariants defined in the usual
way by
\bea
\lefteqn{T_{\mu\nu}(q\cdot p, Q^2) =}\\
& & -T_1(g_{\mu\nu}-\frac{q_\mu q_\nu}{q^2})
 + \frac{T_2}{m_B^2}(p_\mu-q_\mu\frac{q\cdot p}{q^2})
                    (p_\nu-q_\nu\frac{q\cdot p}{q^2})
 -\frac{1}{2m_B^2}T_3i\epsilon_{\mu\nu\alpha\beta}p^\alpha q^\beta\,.
\nonumber
\eea
In our case, because the electromagnetic interactions are
parity--invariant, $T_3=0$. The analytic structure of the scalar
form--factors $T_{1,2}(q\cdot p,Q^2)$ in the $q\cdot p$ complex plane
for fixed $Q^2$ is shown in Fig.1. It has two poles at $q\cdot p =
\pm Q^2/2$, corresponding to the elastic scattering process $\gamma^*B\to
\gamma^*B$ and two cuts for $|q\cdot p| > 1/2[Q^2+(m_B+m_\pi)^2-m_B^2]$.
$T_{1,2}$ are even functions in the variable $q\cdot p$:
\bea
T_{1,2}(-q\cdot p, Q^2) = T_{1,2}(q\cdot p, Q^2)\,.
\eea
The discontinuities of $T_{1,2}$ across the cuts are related to the
hadronic tensor $W_{\mu\nu}$ which describes the inclusive process
$e^-B\to e^-X_b$ (in the approximation that only the b-quark is struck):
\bea
W_{\mu\nu} = \frac{1}{4\pi}\sum_X\langle B|(\bar b\gamma_\mu b)(0)|X\rangle
   \langle X|(\bar b\gamma_\nu b)(0)|B\rangle (2\pi)^4\delta(p_X-p+q)
\eea
which has a decomposition into covariants similar to $T_{\mu\nu}$.
We have
\bea
\mbox{disc}\,T_{\mu\nu}(q,p) &=& 4\pi i W_{\mu\nu}\qquad\mbox{for}\,v\cdot q
<0
\\
\mbox{disc}\,T_{\mu\nu}(q,p) &=& -\mbox{disc}\,T_{\mu\nu}(-q,p)\qquad
   \mbox{for}\,v\cdot q>0\,.
\eea

 To identify the operators which appear in the OPE (1), the usual
method is to take the matrix element of the time-ordered product
between free $b$--quark states. This gives
\bea
\lefteqn{-i\int\mbox{d}^4\!x e^{-iq\cdot x} \langle b(r)|\mbox{T}
(\bar b\gamma_\mu b)(x)(\bar b\gamma_\nu b)(0)|b(r)\rangle =}\\
& &\bar u(r)\gamma_\mu\frac{\rslash-\qslash+m_b}{(r-q)^2-m_b^2}\gamma_\nu u(r)
+\bar u(r)\gamma_\nu\frac{\rslash+\qslash+m_b}{(r+q)^2-m_b^2}\gamma_\mu u(r)
\nonumber
\eea
with $r$ the momentum of the quark. This has to be expanded in powers of
$r$ and the result identified with the matrix elements of a sequence of
operators of the type $\bar b\cdots b$ with an increasing number of
derivatives acting on the fermion field (one for each power of $r$).
At leading order in the coupling $g$, we can replace $r^2$ in the denominators
of (7) by $m_b^2$, since it generates contributions of the form $(iD)^2b$
which equals (in matrix elements) $m_b^2b$ in virtue of the equation of
motion for the $b$ field. The final result is
\bea
\lefteqn{-i\int\mbox{d}^4\!x e^{-iq\cdot x} \mbox{T}(\bar b\gamma_\mu b)(x)
   (\bar b\gamma_\nu b)(0) =}\\
& &-\frac{4}{Q^4}\sum_{n=0}^\infty \left(\frac{2}{Q^2}\right)^{2n}
\lbrace Q^2[g_{\mu\mu_1}g_{\nu\mu_2}-\frac{1}{2}g_{\mu\nu}g_{\mu_1\mu_2}]
+ q_\mu q_{\mu_1} g_{\nu\mu_2}+q_{\mu_1}q_\nu g_{\mu\mu_2}-q_{\mu_1}
q_{\mu_2}g_{\mu\nu}\rbrace\nonumber\\
& & q_{\mu_3}\cdots q_{\mu_{2n+2}}\, P^{\mu_1\mu_2\cdots\mu_{2n+2}}
-\frac{2m_b}{Q^2}g_{\mu\nu}\sum_{n=0}^\infty\left(\frac{2}{Q^2}\right)^{2n}
q_{\mu_1}\cdots q_{\mu_{2n}} R^{\mu_1\mu_2\cdots\mu_{2n}}\,.\nonumber
\eea
Here $P$ and $R$ are defined to be the operators
\bea
P_{\mu_1\mu_2\cdots\mu_n} &=& \frac{1}{n!}\left(
    \bar b\gamma_{\mu_1}(iD)_{\mu_2}\cdots
    (iD)_{\mu_n}b + \mbox{permutations of the indices}\right)\\
Q_{\mu_1\mu_2\cdots\mu_n} &=& \frac{1}{n!}\left(\bar b(iD)_{\mu_1}\cdots
    (iD)_{\mu_n}b + \mbox{permutations of the indices}\right)\,.
\eea
In deriving (8) all operators but the completely symmetric ones have
been neglected, as their matrix elements in an unpolarized hadronic state
vanish.
This is the relation on which all our discussion will be based.
We will take its expectation value between heavy hadron states and the
matrix elements of the operators $P$ and $R$ will be expressed as power
series in $1/m_b$ by using heavy quark effective theory methods. But before
doing so, we pause for a moment to discuss the method by which the
information obtained with the help of the OPE is related to physical
quantities.

    Previous studies \cite{6} of the inclusive semileptonic decays of
hadrons containing one heavy quark have simply identified the
hadronic tensor with the discontinuity of the amplitude $T_{\mu\nu}$
calculated using perturbative QCD. The results thus obtained are highly
singular and must be understood in the sense of duality, when integrated
with an arbitrary smooth function. In this paper we will take a different
approach and will relate the moments of the structure function to the
coefficients of $v\cdot q$ appearing into an expansion of $T_{\mu\nu}$
around the origin in the $v\cdot q$--complex plane with the help of a
dispersion relation \cite{CHM}. In the end the results obtained with the
two methods will be seen to be identical.
 The principle of the method is the
following: for very large $Q^2$, the two cuts on Fig.1 are well separated
and one can use perturbative QCD to reliably compute $T_{\mu\nu}$ around
the point $q\cdot p$=0. In this region each of the
invariant functions $T_{1,2}$ can be expanded in a power series
\beq
T_i(q\cdot p,Q^2) = \sum_{n=0}^\infty T_{in}(Q^2)\left(\frac{2q\cdot p}
{Q^2}\right)^n \,.
\eeq
The coefficients can be expressed as a contour integral
\beq
T_{in}(Q^2) = \frac{1}{2\pi i}\int_C\mbox{d}\!z \frac{T_i(z,Q^2)}{z^{n+1}}
\eeq
where the contour C is a small circle around the origin. It can be deformed
to the shape shown in Fig.1, where it picks contributions from the
discontinuities across the cuts. The contribution of the large circle at
infinity will be neglected. By making use of (5,6), the contour integral
in (12) can be written as
\beq
T_{in}(Q^2) = -2\int_0^1\mbox{d}\! x x^{n-1}W_i(x,Q^2)
\eeq
where $x=Q^2/(2q\cdot p)$ is the Bjorken variable. This is the desired
relation which expresses the moments of the hadronic tensor in terms of
quantities calculable theoretically, the expansion coefficients $T_{in}$.

  We turn now to the evaluation of the transition matrix element
$T_{\mu\nu}$,
by taking the matrix element of the OPE expansion (8) between heavy
meson states. This leads us to the problem of calculating the matrix
elements
\bea
\langle B(v)|P_{\mu_1\mu_2\cdots\mu_n}|B(v)\rangle\,\qquad
\langle B(v)|R_{\mu_1\mu_2\cdots\mu_n}|B(v)\rangle\,.
\eea
They can be expressed as power series in the inverse heavy quark mass
$1/m_b$ by making use of heavy quark effective theory methods. First, the
b-quark field is written as
\beq
b(x) = e^{-im_bv\cdot x}B(x)
\eeq
where
\beq
B(x)=\left [ 1+\frac{i\dslash_\perp}{2m_b} + \frac{1}{4m_b^2}
  \left(v\cdot D\dslash_\perp-\frac{1}{2}\dslash_\perp^2\right ) +
  \cdots\right ]\hb(x)\,.
\eeq
The heavy quark field $\hb$ satisfies $\vslash\hb=\hb$.

  Let us consider for beginning the matrix element of $P$. For simplicity
we exemplify the calculation on the first term in (9), the final result
having to be obtained by symmetrization with respect to all indices.
First we write
\bea
\lefteqn{\langle B(v)|\bar b\gamma_{\mu_1}(iD)_{\mu_2}\cdots(iD)_{\mu_n}b
|B(v)\rangle=}\\
& &\langle B(v)|\bar B\gamma_{\mu_1}(m_bv+iD)_{\mu_2}(m_bv+iD)_{\mu_3}
\cdots(m_bv+iD)_{\mu_n}B|B(v)\rangle\,.\nonumber
\eea
To leading order in the large mass scale this is equal to
\beq
2m_B\,m_b^{n-1}v_{\mu_1}v_{\mu_2}\cdots v_{\mu_n}\,,
\eeq
which is already symmetric in all indices.
Analogously, the matrix element of $R$ can be written as
\bea
\lefteqn{\langle B(v)|\bar B(m_bv+iD)_{\mu_1}(m_bv+iD)_{\mu_2}
\cdots(m_bv+iD)_{\mu_n}B|B(v)\rangle =}\\
& & 2m_B\,m_b^nv_{\mu_1}v_{\mu_2}
\cdots v_{\mu_n} + \mbox{terms suppressed by }1/m_b \,.\nonumber
\eea

  If we insert these expressions for the matrix elements in the equation
(8) for $T_{\mu\nu}$, the following results are obtained for the
invariant transition amplitudes
\bea
T_1(q\cdot p,Q^2) &=& -2\sum_{n=0}^\infty \left(\frac{m_b}{m_B}\right)^{2n+1}
  \left(\frac{2m_Bv\cdot q}{Q^2}\right)^{2n+2}\\
T_2(q\cdot p,Q^2) &=& -8\frac{m_B^2}{Q^2}
  \sum_{n=0}^\infty \left(\frac{m_b}{m_B}\right)^{2n+1}
  \left(\frac{2m_Bv\cdot q}{Q^2}\right)^{2n}\,.
\eea
If we define the more conventional structure functions $F_1(x,Q^2)=W_1(x,Q^2)$
and $F_2(x,Q^2)=q\cdot p/m_B^2 W_2(x,Q^2)$, the following predictions are
obtained for their moments, with the help of Eq.(13)
\bea
& &\int_0^1\mbox{d}\!x x^{n-1}F_1(x,Q^2) =  \left(\frac{m_b}{m_B}\right)^{n-1}
 = \left(1-\frac{\bar\Lambda}{m_B}\right)^{n-1},\qquad n=2,4,6,\dots\\
& &\int_0^1\mbox{d}\!x x^{n-1}F_2(x,Q^2) =  2\left(\frac{m_b}{m_B}\right)^n
 = 2\left(1-\frac{\bar\Lambda}{m_B}\right)^n,\qquad n=1,3,5,\dots
\eea
The evenness property of $T_{1,2}$ (3) prevents us from extracting the
moments (22,23) for other values of $n$ than those indicated. However, these
can be obtained by analytic continuation, that is, by simply assuming that
these relations hold true for all values of $n$.

  Some comments can be made about this result: i) the integral of
$F_1(x,Q^2)$ equals unity (n=1), which corresponds to the parton
interpretation of this function as being the probability for the heavy quark
to be found with a fraction $x$ of the total momentum; ii) as expected,
the Callan-Gross relation $F_L=F_2-2xF_1=0$ is satisfied; iii) the shape of
the distribution function is a simple delta--function
\beq
F_1(x,Q^2) = \delta(x-\frac{m_b}{m_B})
\eeq
which is the naive parton model result, although modified to take into
account the binding energy of the heavy quark in the hadron.

  Let us now go further and examine the effect of retaining more terms
in the matrix elements (17) and (19), suppressed by one or more powers of
$1/m_b$. The correction of order $1/m_b$ to (17) is equal to
\bea
\lefteqn{m_b^{n-1}\sum_{j=2}^nv_{\mu_2}\cdots\vcross_{\mu_j}\cdots v_{\mu_n}
\langle B(v)|\hbbar \gamma_{\mu_1}(iD)_{\mu_j} \hb|B(v)\rangle}\\
& & +
m_b^{n-2}\sum_{j=2}^n v_{\mu_2}\cdots\vcross_{\mu_j}\cdots v_{\mu_n}
\langle B(v)|\hbbar\gamma_{\mu_1}\frac{i\stackrel{\to}{\dslash_\perp}}{2}
\hb|B(v)\rangle\nonumber\\
& & - m_b^{n-2}\sum_{j=2}^n v_{\mu_2}\cdots\vcross_{\mu_j}
\cdots v_{\mu_n}\langle B(v)|\hbbar \frac{i\stackrel{\leftarrow}
{\dslash_\perp}}{2}\gamma_{\mu_1}\hb|B(v)\rangle\nonumber\\
& & + \frac{i}{2}m_b^{n-2}\sum_{j=2}^n\langle B(v)|\int\mbox{d}^4x\mbox{T}
(\hbbar (i\dslash_\perp)^2\hb)(x) (\hbbar\gamma_{\mu_1}\hb|B(v)\rangle
v_{\mu_2}\cdots\vcross_{\mu_j}\cdots v_{\mu_j}\,.\nonumber
\eea

  The matrix element in the first term can be written as
\bea
\langle B(v)|\hbbar \gamma_{\mu_1}(iD)_{\mu_j} \hb|B(v)\rangle =
Av_{\mu_1}v_{\mu_j}
\eea
with
\bea
A=\langle B(v)|\hbbar(iv\cdot D) \hb|B(v)\rangle = 0
\eea
where the equation of motion $iv\cdot D\hb=0$ has been used. The other
three terms have together the form of the $1/m_b$ correction to the
matrix element of the
vector current $\bar b\gamma_{\mu_1}b$, which is known to have no such
corrections. Therefore the entire $1/m_b$ correction to the matrix element
of the $P$ operator vanishes. A similar argument shows that the $1/m_b$
corrections to the matrix elements of $R$ vanish also. From this
follows the conclusion that there are no $1/m_b$ corrections to the
distribution functions of a heavy quark in a heavy hadron.

  In principle the same line of reasoning will give also the $1/m_b^2$
corrections to the distribution functions. This will require the
introduction in the operator product expansion (8) of some new twist
three operators containing one gluon field tensor $F_{\mu\nu}$. Much simpler
it is however to use the result for
$T_{\mu\nu}$ given in the second paper of Ref. \cite{6}. We will show
that the answer obtained with the two methods for the twist-2
contribution is the same. In the next section we discuss the corrections
of order $1/m_b^2$ to the distribution functions of the heavy quarks.
\\[0.5cm]

  {\bf 3.} Blok, Koyrakh, Shifman and Vainshtein have calculated
$T_{\mu\nu}$ up to order $\bar\Lambda^2/m_b^2$ (Eqs.(A1,A2)
in the second paper of Ref.\,\cite{6}). Although
they were interested in a timelike $q$, their expressions are also valid
for a spacelike $q$, the case we need here. The results for $T_{1,2}$ are
\bea
\lefteqn{\frac{1}{2m_B}T_1 = \frac{1}{\Delta}[-v\cdot q+\frac{4m_b}{3}
(K_b+G_b)]}\\
& &+\frac{2m_b}{\Delta^2}\lbrace -\frac{1}{3}G_b[-(v\cdot q)(4m_b-3v\cdot q)
+2[(v\cdot q)^2-q^2]]\nonumber\\
& &+K_b[-(v\cdot q)^2-\frac{2}{3}[(v\cdot q)^2-q^2]]
\rbrace\nonumber\\
& &-\frac{8m_b^2}{3\Delta^3}K_b(v\cdot q)[(v\cdot q)^2-q^2] + (q\to -q)
\nonumber
\eea
and
\bea
\lefteqn{\frac{1}{2m_B}T_2 = \frac{2m_b}{\Delta}[1+\frac{5}{3}
(K_b+G_b)]+\frac{4m_b^2}{3\Delta^2}(7K_b+5G_b)v\cdot q}\nonumber\\
&+ &\frac{16m_b^3}{3\Delta^3}K_b[(v\cdot q)^2-q^2] + (q\to -q)
\eea
with
\beq
\Delta = -2m_bv\cdot q-Q^2+i\epsilon\,.
\eeq
The terms with $q\to -q$ have to be added to account for the necessary
crossing property (3) of $T_{1,2}$. $K_b$ and $G_b$ are defined by
\bea
K_b&=&-\frac{1}{2m_B}\langle B(v)|\hbbar\frac{(iD)^2}{2m_b^2}\hb|B(v)
\rangle=-\frac{\lambda_1}{2m_b^2}=\frac{\mu_\pi^2}{2m_b^2}\\
G_b&=&\frac{1}{2m_B}\langle B(v)|\hbbar\frac{g\sigma_{\alpha\beta}
F^{\alpha\beta}}{4m_b^2}\hb|B(v)\rangle=-\frac{3\lambda_2}{2m_b^2}=
-\frac{\mu_G^2}{2m_b^2}
\eea
and have the approximative values $G_b=-0.0077$ and $K_b=0.01$. We have
shown here the different names by which these quantities are denoted in
the literature \cite{6,M1,M2,R,FN}.

  The expressions (28,29) are the first terms of a power series in the
expansion parameters $\bar\Lambda/m_b$ and $\bar\Lambda^2/\Delta$. Because
we are interested in evaluating these functions around the point $v\cdot q
=0$, the second expansion parameter is actually $\bar\Lambda^2/Q^2$.

  We expand the expressions for $T_{1,2}$ into a power series in
$(2m_Bv\cdot q)/Q^2$ with the result for the coefficient of the $n^{th}$
power
\bea
\lefteqn{T_{1,n} = -2\left(\frac{m_b}{m_B}\right)^{n-1}\lbrace 1+(n-1)
\frac{5G_b+(n+3)K_b}{3}}\nonumber\\
& & +\frac{4}{3}\left(\frac{m_b^2}{Q^2}\right)[4(n+1)G_b+(n^2+3n+4)K_b]\rbrace
\eea
and
\bea
\frac{Q^2}{2m_B^2}T_{2,n}=-4\left(\frac{m_b}{m_B}\right)^{n+1}
\lbrace 1+(n+1)\frac{5G_b+(n+5)K_b}{3}+\frac{4}{3}K_b(n+1)(n+2)\left(
\frac{m_b^2}{Q^2}\right)\rbrace\nonumber\\
\eea
for $n$ even (0,2,4,...) and $T_{i,n}=0$ for $n$ odd.

  According to the discussion in the preceding Section, these coefficients
are directly related to the $(n-1)^{th}$ moment of the respective structure
functions. By performing an inverse Mellin transform or simply by trial, the
functional dependence of $F_{1,2}(x)$, including these corrections, turns out
to be of the form
\bea
F_1(x,Q^2)&=&\left(1+\frac{32}{3}(K_b+G_b)\frac{m_b^2}{Q^2}\right)
\delta(x-\frac{m_b}{m_B})\\
& & - \frac{m_b}{m_B}\left(\frac{5}{3}(K_b+G_b)+
8(K_b+\frac{2}{3}G_b)\frac{m_b^2}{Q^2}\right)
\delta'(x-\frac{m_b}{m_B})\nonumber\\
& & + \frac{1}{3}K_b\left(\frac{m_b}{m_B}\right)^2
\left(1+4\frac{m_b^2}{Q^2}\right)\delta''(x-\frac{m_b}{m_B})\,.\nonumber
\eea
and
\bea
& &\frac{1}{2}\left(\frac{m_B}{m_b}\right)^2xF_2(x,Q^2)=
\left(1+4K_b+\frac{10}{3}G_b+8K_b\frac{m_b^2}{Q^2}\right)
\delta(x-\frac{m_b}{m_B})\\
& & - \frac{m_b}{m_B}\left(3K_b+\frac{5}{3}G_b+8K_b\frac{m_b^2}{Q^2}\right)
\delta'(x-\frac{m_b}{m_B})\nonumber\\
& & + \frac{1}{3}K_b\left(\frac{m_b}{m_B}\right)^2
\left(1+4\frac{m_b^2}{Q^2}\right)\delta''(x-\frac{m_b}{m_B})\,.\nonumber
\eea

  This is the same as what one obtains by directly taking the
discontinuity of (28,29) by using the well-known identity
\beq
\mbox{Im}\,\frac{1}{x+i\epsilon} = -\pi\delta(x)\,.
\eeq

  The Eqs.\,(35,36) are all what QCD predicts about the structure
functions of a heavy quark in a B meson at order $\bar\Lambda^2/m_b^2\,,
\bar\Lambda^2/Q^2$. They have
a rather singular dependence of the Bjorken variable $x$, given by a $\delta$
function and its derivatives. These singularities are similar to what
has been recently obtained for the shape of the electron spectrum near its
end-point in $B\to X_ue\bar\nu$ or for the photon spectrum in
$B\to X_s\gamma$ decays. An interesting feature of these predictions is
that, although the support of the delta functions is restrained to a
single point $x=m_b/m_B$, they do simulate a spreading of the structure
functions from the parton-model prediction. For example, one can compute
the dispersion of the $x$ variable $\sigma_x$ for the structure function
$F_1(x)$:
\bea
\sigma_x^2=\langle x^2\rangle_{F_1} - \langle x\rangle_{F_1}^2=
\frac{m_b^2}{m_B^2}\left(\frac{2}{3}K_b-8(K_b+\frac{4}{3}G_b)\frac{m_b^2}
{Q^2}\right)
\eea
which is nonvanishing and is of the order $\bar\Lambda/m_b$.

  Of course, the physical structure function does not look like a delta
function and derivatives thereof, but has a smooth shape which extends
beyond the parton model boundary\footnote{We assume that the rapid
variations in the physical structure functions due to the resonances
have been smoothed out.}. We will show that the sum of the most singular
terms in the expansions (35,36) to all orders in $\bar\Lambda/m_b$
generates a ``shape function'' which can be considered as an approximation
to the physical structure function. The argument goes in close
analogy to the analysis of Neubert of the photon spectrum in $B\to
X_s\gamma$ decays. However, the details of the derivation are somewhat
different, as appropriate to the problem at hand.

  We will consider in the following the scaling limit $Q^2\to \infty$
when the distribution functions depend only on the Bjorken variable $x$.
Note that these are not the distribution functions which would be
measured in a hypothetical electroproduction experiment on B mesons,
since the relevant matrix elements are evaluated at the heavy quark scale
and the observable distribution functions would have to be obtained
by evolution to the large momentum scale $Q^2$. However, we will have
nothing to say about this evolution and therefore all our results refer
exclusively
to the ``boundary data'' for the distribution functions at the scale
$\mu^2=m_b^2$.

  In this limit there is only one independent structure function, since
the two functions $F_1(x)$ and $F_2(x)$ are connected by the
Callan-Gross relation $F_2=2xF_1$. One can easily convince himself that
the predictions (35,36) satisfy this equality in the limit $Q^2\to\infty$.
It is only violated at first order in $\alpha_s$. Therefore, we will
proceed further with $F_1(x)$ and call it simply ``the distribution
function''.

  The physical distribution function in the scaling limit $F_1(x)$
can be represented, when convoluted with an arbitrary smooth function,
as the sum of a delta function and its derivatives:
\beq
F_1(x)=\sum_{i=0}^\infty(-1)^i\frac{M_i}{i!}\delta^{(i)}(x-\frac{m_b}{m_B})
\eeq
where the coefficients $M_i$ are the moments of $F_1(x)$
\beq
M_i=\int_0^1\mbox{d}\!x (x-\frac{m_b}{m_B})^iF_1(x)\,.
\eeq
The first few moments $M_i$ can be extracted from Eq.(35)
\bea
M_0&=&1\\
M_1&=&\frac{5}{3}\frac{m_b}{m_B}(K_b+G_b)\\
M_2&=&\frac{2}{3}\frac{m_b^2}{m_B^2}K_b
\eea
Each of the moments $M_i$ can be expanded in a power series in $\bar\Lambda
/m_b$ and it is reasonable to assume that keeping only the leading term
in this expansion provides a good first approximation to the description
of $F_1(x)$. This is equivalent to keeping only the leading singularities in
(38) (the smallest power of $\bar\Lambda/m_b$ corresponding to a derivative
of a given order of the $\delta$ function is the same as the highest order of
the derivative corresponding to a given order in $\bar\Lambda/m_b$).

  In the scaling limit $Q^2\to\infty$ only the operator $P$ in (9) gives a
nonvanishing contribution to the distribution functions, since it is the
only one which
contains twist-2 operators. The operator $R$ and any other operators
involving the gluon field tensor give contributions suppressed by one
or more powers of $Q^2$. The matrix elements of $P$ have the form
\bea
\langle B(v)|P_{\mu_1\mu_2\cdots\mu_n}|B(v)\rangle =
Cv_{\mu_1}v_{\mu_2}\cdots v_{\mu_n} +\,
\mbox{terms containing $g_{\mu\nu}$'s}
\eea
We are only interested in the part proportional to $v$'s because the
terms with $g_{\mu\nu}$ vanish in the scaling limit. Let us try to extract
as much information as possible about the constant $C$ to all orders in
$1/m_b$. It has an expansion of the form
\beq
C=2m_B m_b^{n-1}\left( 1+\epsilon C_1+\epsilon^2 C_2+\cdots\right)
\eeq
with
\beq
\epsilon = \frac{\bar\Lambda}{m_b}
\eeq
A direct calculation by making use of the methods exposed in Section 2 gives
\bea
\epsilon C_1&=&0\\
\epsilon^2 C_2&=&\frac{2}{3}K_b\frac{(n-1)(n-2)}{2}+\frac{2}{3}(K_b+G_b)
(n-1)+(K_b+G_b)(n-1)
\eea
The first relation expresses the vanishing of the $1/m_b$ corrections to
the structure functions, noted in the preceding Section. The three terms
in the second relation have respectively, the following origin: i) the
first appears from the matrix element $\langle B(v)|(iD)_\mu(iD)_\nu|
B(v)\rangle$ resulting from the expansion of (17);
ii) the second appears from the correction to the $B$ field (16) at order
$1/m_b$ and iii) the third appears from the insertion of the $1/m_b$
term in the HQET Lagrangian.

  If the expression (44) together with (45,47,48) are introduced into
Eq.(8) and the result for $T_1$ expanded in a power series in $(2m_Bv\cdot
q)/Q^2$, the coefficient of the $n^{th}$ power which is obtained coincides
with Eq.(33) (in the limit $Q^2\to\infty$). This establishes the equivalence
of the two approaches to this order in $\bar\Lambda/m_b$.

 The constant $C$ is directly related to the $(n-1)^{th}$ moment of the
structure function $F_1(x)$, as can be seen upon inserting (45) into the
general formula (8):
\bea
\int_0^1\mbox{d}\! xx^{n-1}F_1(x)=\left(\frac{m_b}{m_B}\right)^{n-1}
\left(1+\epsilon C_1+\epsilon^2 C_2+\cdots\right)
\eea

  For a given order in $\epsilon$ (say, $\epsilon^k$), the most singular
terms in $F_1$ correspond to those terms in the coefficient $C_k$ which
have the highest power of $n$. For example, the term with the highest power
of $n$ in $C_2$ is $\frac{1}{3}K_bn^2$ and it corresponds to the contribution
$\frac{1}{3}K_b\left(\frac{m_b}{m_B}\right)^2\delta''(x-\frac{m_b}{m_B})$
in $F_1(x)$, Eq.(35). As said before, this term originated from simply
expanding the brackets in
\beq
\langle B(v)|\bar B (m_bv+iD)_{\mu_1}(m_bv+iD)_{\mu_2}\cdots
(m_bv+iD)_{\mu_n}B |B(v)\rangle\,,
\eeq
replacing $B$ by the HQET field $\hb$ and forgetting about lagrangian
insertions. We will show now by a simple counting argument that this holds
true generally and that the leading terms in $n$ are correctly reproduced by
retaining only these contributions, to all orders in $\epsilon$.

  The contribution to $C_k$ of these terms can be written as
\bea
\frac{1}{2m_Bm^{k}}\sum\langle B(v)|\hbbar\gamma_{\mu_1}(iD)_{\alpha_1}
(iD)_{\alpha_2}\cdots(iD)_{\alpha_k}\hb |B(v)\rangle v_{\alpha_{k+1}}
\cdots v_{\alpha_n}
\eea
where the sum runs over all possible ways to choose the $k$ derivatives
out of a total of $(n-1)$ possibilities, and contains thus
$\frac{(n-1)!}{k!(n-k-1)!}$ terms.

  Let us consider now a contribution to $C_k$ arising from corrections
to the $B$ field. We write the expansion (16) of $B$ in powers of $1/m_b$
in a compact form as
\beq
B(x)=\sum_{n=0}^\infty\frac{B_n}{m_b^n}\hb(x)
\eeq
with $B_0=1, B_1=(i\dslash)/2$, etc. Then the desired contribution reads
\bea
\frac{1}{2m_Bm^{k}}\sum_{0\leq i+j\leq k-1}\langle B(v)|\hbbar\bar B_i
\gamma_{\mu_1}(iD)_{\alpha_1}(iD)_{\alpha_2}\cdots(iD)_{\alpha_{k-i-j}}
B_j\hb |B(v)\rangle v_{\alpha_{k-i-j+1}}\cdots v_{\alpha_n}\nonumber\\
\eea
and the sum contains
\bea
\sum_{i=0}^{k-1}\sum_{j=0}^{k-i-1}\frac{(n-1)!}{(k-i-j)!(n-k+i+j-1)!}=
(n-1)!\sum_{r=0}^{k-1}\frac{r+1}{(k-r)!(n-k+r-1)!}
\eea
terms. We can omit the terms with $i+j=k$ because this is a correction to
the matrix element of the vector current which is known to vanish.
We have included here also the preceding case as the term with
$i=j=0$ respectively $r=i+j=0$. Assuming all the matrix elements in
(53) to have about the same order of magnitude, one can see that the
total contribution from the matrix elements without corrections to the $B$
field ($r=0$) scales like $n^k$ for a given order $k$ in the $\epsilon$
expansion. The next contributions, containing a suppression by $1/m_b$ from a
correction to the heavy quark field ($r=1$), scale as $n^{k-1}$, and so on.
The matrix elements containing insertions of higher-dimensional terms
in the HQET lagrangian can be shown in a similar fashion to give only
subleading contributions in $n$. This proves the above-mentioned assertion
that the most singular terms in $F_1(x)$ are correctly reproduced by
replacing
\bea
& &\langle B(v)|P_{\mu_1\mu_2\cdots\mu_n}|B(v)\rangle \to\\
& &v_{\mu_1}\langle B(v)|\hbbar (m_bv+iD)_{\mu_2}(m_bv+iD)_{\mu_3}\cdots
(m_bv+iD)_{\mu_n}\hb|B(v)\rangle\nonumber\\
& & +\,\mbox{symmetrization w.r.t. all indices}\nonumber
\eea

  We define the dimensionless constants $a_n\simeq 1$ through
\beq
\frac{1}{2m_B}\langle B(v)|\hbbar (iD)_{\mu_1}(iD)_{\mu_2}\cdots(iD)_{\mu_n}
\hb|B(v)\rangle=a_n\bar\Lambda^n v_{\mu_1}v_{\mu_2}\cdots v_{\mu_n}+\cdots
\,.
\eeq
The first few $a_i$ are
\bea
a_0&=&1\\
\epsilon a_1&=&0\\
\epsilon^2a_2&=&\frac{2}{3}K_b\,.
\eea
In terms of these constants the matrix element (55) can be written as
\bea
\lefteqn{v_{\mu_1}\langle B(v)|\hbbar (m_bv+iD)_{\mu_2}(m_bv+iD)_{\mu_3}\cdots
(m_bv+iD)_{\mu_n}\hb|B(v)\rangle}\\
&+&\,\mbox{symmetrization w.r.t. all indices}
=2m_Bm_b^{n-1}\sum_{i=0}^{n-1}\frac{(n-1)!}{i!(n-i-1)!}a_i
\frac{\bar\Lambda^i}{m_b^i}v_{\mu_1}v_{\mu_2}\cdots v_{\mu_n}\,.\nonumber
\eea
Inserting this in Eq.(8) gives for the moments of $F_1(x)$ (only the
part containing the leading singularities, the so-called ``shape function'')
\bea
\int_0^1\mbox{d}\! xx^{n-1}F_1(x)=\left(\frac{m_b}{m_B}\right)^{n-1}
\sum_{i=0}^{n-1}\frac{(n-1)!}{i!(n-i-1)!}a_i\epsilon^i\,.
\eea

  The moments $M_i$ with respect to the parton-model point $x=m_b/m_B$ which
appear in (39) can be obtained from the above ones by a simple expansion
\bea
\lefteqn{M_{n-1}=\int_0^1\mbox{d}\! x(x-\frac{m_b}{m_B})^{n-1}F_1(x)=
\sum_{i=0}^{n-1}\frac{(n-1)!}{i!(n-i-1)!}\left(-\frac{m_b}{m_B}
\right)^{n-i-1}\int_0^1\mbox{d}\! xx^iF_1(x)}\nonumber\\
&=&\left(\frac{m_b}{m_B}\right)^{n-1}
\sum_{i=0}^{n-1}\sum_{j=0}^i(-1)^{n-i-1}\frac{(n-1)!}{j!(n-i-1)!(i-j)!}a_j
\epsilon^j = \left(\frac{m_b}{m_B}\right)^{n-1}\epsilon^{n-1}a_{n-1}\,,
\nonumber\\
\eea
where the final result has been obtained by interchanging the summation
order. It has been shown \cite{M1,M2,R} that the line shape in the
$B\to X_s\gamma$ decays
and the end-point electron spectrum in the semileptonic $B\to X_u$ decays
are essentially determined by one common universal function whose moments
are related to the nonperturbative matrix elements $a_i$ in a fashion
similar to (62). Therefore the result (62) is remarkable, as it shows that
the same universal function gives also the heavy quark structure functions
for electroproduction on a B meson.

  Most probably, these structure functions will never be measured and
therefore our results are of little practical value. Nevertheless, we
believe that they are interesting in themselves because they show how
completely
different phenomena involving heavy quarks are governed by similar
distribution functions. Finally, we mention the related problem, of
considerable experimental interest, of the fragmentation functions for
heavy quarks into heavy hadrons, for which one cannot make use of the
powerful formalism of the operator product expansion. One can hope that
similar methods could be used to study the shape of these functions,
and set constraints on the possible structure of their nonperturbative
contributions.\\[0.5cm]

  {\bf 4.} In this Section we derive two new sum rules for the
nonperturbative functions which describe the form-factors for
heavy-to-heavy transitions. These sum rules are deduced for the process
$\gamma^*+B\to X_b$, with $X_b$ an arbitrary excited state of the B
meson, but they can be extended by using heavy quark symmetry for the
case of greater interest $B\to X_c+W$, with $X_c$ an excited state of
the $D$ meson.

   We will relax the assumption $Q^2\to\infty$ and keep $Q^2$ arbitrary
(but still much larger than $\Lambda_{QCD})$.
Let us examine the form of the physical structure functions $F_{1,2}(x,Q^2)$.
In the neighbourhood of $x=1$ (near the end of the cuts) they have a rapid
variation, with spikes
corresponding to each resonance $X_b$ superposed over a smooth background
due to two- and many-body intermediate states, such as $B\pi$. The
contribution of a particular resonance of mass $m_X$ to $W_{\mu\nu}$ can
be obtained from (4) to be given by
\bea
\lefteqn{W_{\mu\nu}^{(X)}(p\cdot q,Q^2) = \frac{1}{4}
 \delta [p\cdot q-\frac{1}{2}(m_B^2-m_X^2-Q^2)]\times}\\
& & \langle B(v)|(\bar b\gamma_\mu b)(0)|X_b(v')\rangle
 \langle X_b(v')|(\bar b\gamma_\nu b)(0)|B(v)\rangle\nonumber
\eea
where the velocity $v'$ is given by
\bea
v\cdot v' = \frac{m_B^2+m_X^2+Q^2}{2m_B m_X}\,.
\eea
The matrix elements can be expanded in powers of $1/m_b$ by using heavy
quark effective theory methods and the resulting structure functions
$F_{1,2}$ are expressed in terms of the usual Bjorken variable $x$ by using
\bea
\delta [p\cdot q-\frac{1}{2}(m_B^2-m_X^2-Q^2)] =
\frac{2x_X^2}{Q^2}\delta(x-x_X)
\eea
with
\beq
x_X = \frac{Q^2}{Q^2-m_B^2+m_X^2}\,.
\eeq

  The key ingredient which makes possible writing a sum rule is the
positivity property of $F_{1,2}$ \cite{IZ}, which expresses the fact that each
individual intermediate state in (4) contributes a positive quantity.
Equating the OPE result for the moments (33,34) with the moments of the
structure functions written as sums over intermediate states gives an
upper bound on any partial sum of the latter.

  On the tip of the cut, $m_X=m_B$ and $v\cdot v'=1+Q^2/(2m_B^2)$. The
velocity transfer $v\cdot v'$ decreases as we move along
the cut until it reaches a minimum $(v\cdot v')_{min} = \sqrt{1+Q^2/m_B^2}$
for $m_X=m_B\sqrt{1+Q^2/m_B^2}$ after which it starts increasing.
Therefore the various intermediate states will appear in the sum rule
with a different velocity transfer $v\cdot v'$.

{}From (63) we get
\bea
\lefteqn{W_{\mu\nu}^{X}(x,Q^2) = \frac{Q^2}{2(Q^2+m_X^2-m_B^2)^2}\delta(x-x_X)
}\\
& & \langle B(v)|(\bar b\gamma_\mu b)(0)|X_b(v')\rangle
 \langle X_b(v')|(\bar b\gamma_\nu b)(0)|B(v)\rangle\nonumber
\eea
or, in explicit form for the s- and p-wave mesons
\bea
F_1^i(x,Q^2) = \frac{Q^2}{2(Q^2+m_{X_i}^2-m_B^2)^2}\delta(x-x_{X_i})R^i
\eea
with
\begin{itemize}
\item B meson $R=0$.
\item B$^*$ meson $R=m_Bm_{B^*}(\omega^2-1)|\xi+\frac{1}{m_b}
   (2\chi_1-2(\omega-1)\chi_2+4\chi_3+\bar\Lambda\xi-\xi_3)|^2$.
\item B$_0$ meson ($s_\ell^{\pi_\ell}=1/2^+, s^\pi=0^+$) $R=0$.
\item B$_1$ meson ($s_\ell^{\pi_\ell}=1/2^+, s^\pi=1^+$) $R=
  \frac{1}{3}m_Bm_{B_1}(\omega-1)^2|\xi_{1/2}|^2$.
\item B$'_1$ meson ($s_\ell^{\pi_\ell}=3/2^+, s^\pi=1^+$) $R=
  \frac{1}{6}m_Bm_{B'_1}(\omega^2-1)^2|\xi_{3/2}|^2$.
\item B$_2$ meson ($s_\ell^{\pi_\ell}=3/2^+, s^\pi=2^+$) $R=
  \frac{1}{2}m_Bm_{B_2}(\omega^2-1)^2|\xi_{3/2}|^2$.
\end{itemize}
and for $xF_2(x,Q^2)$
\bea
xF_2^i(x,Q^2) = \frac{Q^4}{(Q^2+m_{X_i}^2-m_B^2)^2}\delta(x-x_{X_i})S^i
\eea
with
\begin{itemize}
\item B meson $S=|\xi+\frac{2}{m_b}(\chi_1-2(\omega-1)\chi_2+6\chi_3)|^2$.
\item B$^*$ meson $S=\frac{m_{B^*}}{4m_B}\left(-1-\frac{m_B^2}{m_{B^*}^2}
  +2\frac{m_B}{m_{B^*}}\omega\right)|\xi+\frac{1}{m_b}
   (2\chi_1-2(\omega-1)\chi_2+4\chi_3+\bar\Lambda\xi-\xi_3)|^2$.
\item B$_0$ meson ($s_\ell^{\pi_\ell}=1/2^+, s^\pi=0^+$) $S=0$.
\item B$_1$ meson ($s_\ell^{\pi_\ell}=1/2^+, s^\pi=1^+$) $S=
  \frac{1}{6}(\omega-1)|\xi_{1/2}|^2$.
\item B$'_1$ meson ($s_\ell^{\pi_\ell}=3/2^+, s^\pi=1^+$) $S=
  \frac{m_{B'_1}}{8m_B}\left(1+\frac{m_B^2}{m_{B'_1}^2}
  +\frac{2m_B}{3m_{B'_1}}(\omega+4)\right)(\omega^2-1)|\xi_{3/2}|^2$.
\item B$_2$ meson ($s_\ell^{\pi_\ell}=3/2^+, s^\pi=2^+$) $S=
  \frac{m_{B_2}}{8m_B}\left(-1-\frac{m_B^2}{m_{B_2}^2}
  +2\frac{m_B}{m_{B_2}}\omega\right)(\omega^2-1)|\xi_{3/2}|^2$.
\end{itemize}

Here $\omega$ stands for the velocity change $v\cdot v'$.
$\xi(\omega),\xi_{1/2,3/2}(\omega)$ are the Isgur-Wise functions which
describe the form-factors of the transitions from the respective multiplets
to the lowest--lying one\footnote{$\xi_{1/2,3/2}$ are related to the
form-factors used in \cite{a} by $\xi_{1/2}=2\sqrt{3}\tau_{1/2}$ and
$\xi_{3/2}=\sqrt{3}\tau_{3/2}$.}.
For the s-wave multiplet we have included also the contribution from
the subleading form-factors $\chi_{1,2,3}, \xi_3$, which are defined as in
\cite{Luke,FN}. For example, if we assume for the moment that the
multiplets are degenerate, their contribution to $F_{1,2}$ takes the form
\bea
\lefteqn{F_1^{res}(x,Q^2) = \frac{1+\omega}{2}
|\xi+\frac{1}{m_b}(2\chi_1-2(\omega-1)\chi_2+4\chi_3+\bar\Lambda\xi-
\xi_3)|^2\delta(x-1)}\nonumber\\& & +
\frac{Q^2[Q^2+(m_{B_1}-m_B)^2]}{6[Q^2+(m_{B_1}^2-m_B^2)]^2}(\omega-1)
   |\xi_{1/2}|^2 \delta(x-x_{B_1})\\& &+
\frac{Q^2[Q^2+(m_{B_2}-m_B)^2]}{3[Q^2+(m_{B_2}^2-m_B^2)]^2}(\omega-1)
   (1+\omega)^2|\xi_{3/2}|^2\delta(x-x_{B_2})+\cdots\nonumber
\eea
and
\bea
\lefteqn{xF_2^{res}(x,Q^2) = \frac{1+\omega}{2}
|\xi|^2\delta(x-1)}\nonumber\\& & +
\frac{Q^4}{6[Q^2+(m_{B_1}^2-m_B^2)]^2}(\omega-1)|\xi_{1/2}|^2 \delta(x-x_{B_1})
\\& &+
\frac{Q^4}{3[Q^2+(m_{B_2}^2-m_B^2)]^2}(\omega-1)
   (1+\omega)^2|\xi_{3/2}|^2\delta(x-x_{B_2})+\cdots\nonumber
\eea

However, since we work to an arbitrary order in $1/m_b$, we will consider
henceforth the masses of the two members of each multiplet as different.

   The sum rules result when the moments of $F_{1,2}$ expressed as a sum over
resonances are equated with the QCD prediction
\bea
\lefteqn{\int_0^1\mbox{d}\!x x^{n-1}F_1(x,Q^2) = \left(\frac{m_b}{m_B}
\right)^{n-1}\lbrace 1+(n-1)\frac{5G_b+(n+3)K_b}{3}}\nonumber\\
& & +\frac{4}{3}\left(\frac{m_b^2}{Q^2}\right)
[4(n+1)G_b+(n^2+3n+4)K_b] + \cdots\rbrace
\eea
and
\bea
\int_0^1\mbox{d}\!x x^nF_2(x,Q^2) =\left(\frac{m_b}{m_B}\right)^{n+1}
\lbrace 1+(n+1)\frac{5G_b+(n+5)K_b}{3}\nonumber\\
+\frac{4}{3}K_b(n+1)(n+2)\left(
\frac{m_b^2}{Q^2}\right)+\cdots\rbrace
\eea

  At this point a serious problem emerges which might diminish the
usefulness of the sum rules. The most interesting region where one would
like to apply these sum rules is in the vicinity of the equal-velocity
point $v\cdot v'=1$, which corresponds to $Q^2=0$. But, as one can see from
(72,73), the $1/m_b^2$ corrections become very large in this limit, and
therefore one can trust no longer the expansion in powers of $1/m_b$.
This happens because the expansion parameter in this problem is
$\bar\Lambda/Q$. For example, if we require as a limit of applicability
that $\bar\Lambda/Q=0.5$, this would correspond to $v\cdot v'\simeq 1.018$
on the tip of the cut.

  There exists yet another way to apply the sum rules, which allows one to
consider practically any value of $v\cdot v'$. This can be done if
the $b$ quark mass is considered as a free parameter. Then one can proceed in
the following way: i) the ratio $Q^2/m_B^2$
must be kept fixed to a value given by the velocity change $v\cdot v'$
one wishes to investigate, according to (64); ii) $m_B$ and $Q$ must be sent
simultaneously to infinity (compared to $\bar\Lambda$, which is of course
fixed); iii) the result is expanded in inverse powers of $m_B$. This
procedure makes sure that the higher order corrections in (72,73) stay
smaller than the leading term.

   In the limiting case when $Q^2,m_B\to\infty$ with $Q^2/m_B^2=2(\omega-1)$
the two sum rules for $F_1$ and $xF_2$ reduce to the usual Bjorken sum rule
\cite{a}:
\bea
1 = \frac{1+\omega}{2}|\xi|^2 + \frac{1}{6}(\omega-1)
   |\xi_{1/2}|^2 + \frac{1}{3}(\omega-1)(1+\omega)^2|\xi_{3/2}|^2+\cdots\,.
\eea

  For each $n$, a different sum rule is obtained. Higher values of $n$
suppress the contributions of the excited states, since $x_X$ is smaller
for these. We note that one cannot write one sum rule for each power of
$1/m_b$, since the various contributions will not contribute in general
with the same sign, so we cannot set constraints on the subleading
form-factors appearing to a given order in the heavy mass expansion. Rather,
the sum rules can be used to test the consistency of a set of calculated
form-factors: the total contribution of a number of resonant states, calculated
to a given order in $1/m_b$, must not exceed the QCD prediction for the
sum of all excited states, calculated to the same order in $1/m_b$.

  {\em Note added.} After completing this work, the paper \cite{new1}
appeared, which extends the results in \cite{M1,M2,F,R} to cases when the
final quark mass is nonvanishing. Also, a different derivation of the
Bjorken sum rule has been recently given in \cite{new2}.\\[0.5cm]

   {\em Acknowledgements.} I am grateful to A.Czarnecki, C.Diaconu,
Z.Gagyi-Palffy, J.K\"orner, M.Neubert and K.Schilcher for discussions and
encouragements.\\[1cm]

\section*{Figure Caption}

{\bf Fig.1} The analytical structure of the forward Compton scattering
amplitude $T_{\mu\nu}$ as a function of $p\cdot q$ for given $Q^2$.


\begin{thebibliography}{12}
\bibitem{6}
   I.I.Bigi, M.Shifman, N.G.Uraltsev and A.I.Vainshtein,
   {\em Phys.Rev.Lett.}{\bf 71} (1993) 496\\
   B.Blok, L.Koyrakh, M.Shifman and A.I.Vainshtein, NSF-ITP-93-68,
   TPI-MINN-93/33-T, UMN-TH-1208/93, July 1993\\
   A.Manohar and M.Wise, UCSD/PTH 93-14, CALT-68-1883, August 1993\\
   T.Mannel, IKDA 93/26, September 1993
\bibitem{FLS} A.F.Falk, M.Luke and M.J.Savage, SLAC preprint SLAC-PUB-6317
   (1993)
\bibitem{CGG} J.Chay, H.Georgi and B.Grinstein, {\em Phys.Lett.}{\bf B247}
   (1990) 399
\bibitem{M1} M.Neubert, CERN-TH.7087/93, November 1993
\bibitem{M2} M.Neubert, CERN-TH.7113/93, December 1993
\bibitem{F} A.F.Falk, E.Jenkins, A.V.Manohar and M.B.Wise, UCSD-PTH-93-38,
   December 1993
\bibitem{R} I.I.Bigi, M.A.Shifman, N.G.Uraltsev and A.I.Vainshtein,
   TPI-MINN-93/60-T, UMN-TH-1231-93, CERN-TH.7129/93, December 1993
\bibitem{BJ} J.D.Bjorken, SLAC preprint SLAC-PUB-5278 (1990)\\
   J.D.Bjorken, I.Dunietz and J.Taron, {\em Nucl.Phys.}{\bf B 371}(1992)111
\bibitem{JR} R.L.Jaffe and L.Randall, {\em Nucl.Phys.}{\bf B412}(1994)79
\bibitem{CHM} N.Christ, B.Hasslacher and A.H.Mueller, {\em Phys.Rev.}
   {\bf D6}(1972)3543
\bibitem{FN} A.F.Falk and M.Neubert, {\em Phys.Rev.}{\bf D47}(1993)2965
\bibitem{IZ} C.Itzykson and J.Zuber, {\em Quantum Field Theory},
   Mc Graw-Hill, N.Y. 1980
\bibitem{a} N.Isgur and M.B.Wise, {\em Phys.Rev.}{\bf D43}(1991)819\\
   N.Isgur, M.B.Wise and M.Youssefmir, {\em Phys.Lett.}{\bf B254}(1991)215
\\
   W.Roberts, {\em Nucl.Phys.}{\bf B389}(1993)549
\bibitem{Luke} M.E.Luke, {\em Phys.Lett.}{\bf B252}(1990)447
\bibitem{new1} T.Mannel and M.Neubert, CERN-TH.7156/94, February 1994
\bibitem{new2} T.D.Cohen and J.Milana, UMPP \#94-086, February 1994
\end{thebibliography}
\end{document}